\documentstyle [11pt, twoside] {article}
\newcommand{\bye}{\end{document}}
\newcommand{\be}{\begin{equation}}
\newcommand{\gsim}{\lower.7ex\hbox{$\;\stackrel{\textstyle>}{\sim}\;$}}
\newcommand{\lsim}{\lower.7ex\hbox{$\;\stackrel{\textstyle<}{\sim}\;$}}
\def\NPB#1#2#3{Nucl. Phys. B {\bf#1} (19#2) #3}
\def\PLB#1#2#3{Phys. Lett. B {\bf#1} (19#2) #3}

\def\PRD#1#2#3{Phys. Rev. D {\bf#1} (19#2) #3}
\def\PRL#1#2#3{Phys. Rev. Lett. {\bf#1} (19#2) #3}

\newcommand{\nummer}[1]{\baselineskip=14pt\hskip 12 cm #1 \par}
\newcommand{\datum}[1]{\hskip 12 cm #1}
\newcommand{\titel}[1]{\Large \vskip 5 true cm\begin{center}#1\end{center}
            \normalsize\vskip 1.0 true cm}
\newcommand{\autor}[1]{\normalsize\begin{center}#1\end{center}\vskip 0.5cm}
\newcommand{\adresse}[1]{\begin{center}#1\end{center}\vskip 3 true cm}
\renewcommand{\abstract}[1]{\hfil\parbox{15.4 true cm}
{\large ABSTRACT: \normalsize #1} \hfil \vfil \Large
\normalsize\eject}
\begin{document}

\nummer{WU B 97-23}
\datum{July  \quad 1997}

\titel
{{\Large \bf Massive Neutrinos and Lepton Mixing\\
 in Unified  Theories}\footnote{Talk presented at the XVI
 International Workshop  on Weak Interactions and Neutrinos,
 Capri, Italy, June 22-28, 1997.}}

\autor{{\large Yoav Achiman}~\footnote{e-mail: achiman@wpts0.physik.uni-wuppertal.de}}
\adresse{Department of Physics \\
         University of Wuppertal \\
         Gau\ss{}str. 20, D-42097 Wuppertal \\
         Germany}
\abstract{
The recent GUT ($\times$ SUSY) models which can predict the neutrino properties are reviewed.} 
\vfill\eject

The fact that talks about neutrinos dominated this workshop, shows the growing 
popularity of this subject. Neutrino physics  will be probably the first field to teach us
about the physics beyond the standard model (SM). It is interesting therefore to know
what are the predictions  for the neutrino-properties in  possible extensions of the SM. 
The most natural ones are obviously the Grand Unified Theories (GUTs) and
Supersymmetric-GUTs (SUSY-GUTs) and I will limit myself in this talk to  
those theories.  ``Low-energy'' anti-GUT models were 
discussed in the talk of Valle~\cite{val}. 
 
{\bf Neutrinos and the problem of fermionic masses}\\
The masses of the charged fermions are arbitrary in the SM and the neutrinos remain 
massless\footnote{as the SM has no RH neutrinos and the accidental B-L symmetry protects the 
masslessness.}. 
The SM must be therefore extended to be able to account for the fermionic masses. The
extension into GUTs is in particular interesting as they lead to relations between the
masses (but those are not enough and one needs on top of that also a family symmetry).
 
In contrast with the charged fermions the neutrino-masses and lepton-mixing are unknown
phenomenologically. The subject of this talk is to review GUT models which can predict the
neutrino sector. The idea is, to look for theories which can use known information about 
charged fermions to fix the neutrino Dirac mass matrix $M_\nu^D$ as well as the RH
neutrino one $M_\nu^R$. In this way the light neutrino mass matrix is obtained using the
see-saw mechanism as   
\begin{equation}
M_\nu^{light} = -M_\nu^D {M_\nu^R}^{-1} {M_\nu^D}^T   \quad .
\end{equation}
 
This is the most natural way to obtain light neutrino masses when the model involves a large
mass scale $M_{\nu_R}$, like in the GUTs.  
 
{\bf Why GUTs ($\times$ SUSY)?} \\
There is a large list of indications for GUTs:
\begin{itemize}
\item{The unification of the gauge coupling constants.}
\item{The Yukawa unification: $m_\tau(M_{GUT}) \simeq m_b(M_{GUT})$\\and other relations between the masses which are very useful for  predicting  the
neutrino  properties.}
\item{The dynamical electroweak symmetry breaking obtained when the soft SUSY breaking is
universal at $M_{GUT}$.}
\end{itemize}
ALL AT THE SAME SCALE $M_{GUT} \simeq 10^{16} GeV$~, also
\begin{itemize}
\item {The high scale allows for the see-saw   mechanism.}
\item{GUTs can naturally explain the baryon asymmetry  directly or induced via
leptogenesis.}
\item{Modern cosmology requires GUTs.}
\item{Superstring theories have GUTs as a possible intermediate effective   theory.}
\end{itemize}
\begin{center}
GUTS CANOT EXPLAIN EVERYTHING BUT THEY ARE VERY GOOD EFFECTIVE
THEORIES AT $M_{GUT}  \simeq 10^{16} GeV$
\end{center}
{\bf What are the predictive GUTs for the neutrino sector?}\\
Min. SU(5) ($\times$ SUSY)  -- has no RH  neutrinos and the vanishing neutrino masses
are protected by B-L. Small masses can be obtained via gravitations affects but no real
prediction for the leptonic mixing emerges. We will discuss here therefore SO(10) and 
$E_6$ only. 
 
{\bf What is the mass scale of the RH neutrinos (needed for the see-saw mechanism)?}\\
The relevant scale for MSW solution for the Solar Neutrino Puzzle (SNP)~\cite{snp}
and neutrinos as dark matter is $M_{\nu_R} \simeq10^{12} GeV \ .$~\cite{val} \\
The interesting point about this scale is that it is also the scale of other phenomenae like 
\begin{itemize}
\item{the B-L  violation}
\item{the breaking of the Peccei-Quinn symmetry}
\item{the leptogenesis which induces the baryon asymmetry}
\item{the SUSY breaking in the hidden space}
\end{itemize} 
SUSY-GUTs however do not allow for an intermediate symmetry. Hence the natural scale for
SUSY-GUTs is $M_{\nu_R} \simeq M_{GUT}$\ . Such a scale can be used for the vacuum 
oscillation solution to the SNP  because 
$m_c^2 / M_{GUT}  \approx 10^{-6}  eV$\ . 
But most SUSY-GUTs models for the neutrinos use still $M_{\nu_R}  \simeq 10^{12} GeV$\ and
this requires in general fine tuning\footnote{Note, however, that 
$\frac{(M_{GUT})^2}{M_{Planck}} \approx 10^{13} GeV$
what suggests that non-renormalizable contributions due to an underlying theory above the 
GUT (superstrings?) may generate such a scale.}.

It is interesting to note here that allowing for an intermediate scale around  $M_I \simeq
10^{12} GeV$, one can break L-R symmetric GUTs, like SO(10) and $E_6$, in two
steps and obtain gauge and Yukawa unification even without SUSY. 
Also, before going to discuss explicit models let me make one more general remark.
It is relatively ``simple'' to fix $M_\nu^D$  using GUT relations, but simple GUTs cannot 
say a thing about  $M_\nu^R$ . Many models use therefore ad-hoc assumptions about
$M_\nu^R$~. E.g. taking it proportional to a unite matrix or to a diagonal one with a given
hierarchy. However,TO PREDICT THE NEUTRINO PROPERTIES, THE MODEL
  MUST FIX THE RH NEUTRINO MASS MATRIX $M_\nu^R$ AND NOT ONLY THE
  DIRAC ONE $M_\nu^D$. 
\vfill\eject
{\bf Examples for masses models for the charged fermions and their extension into
the neutrino sector:}
 
{\bf Texture Zeros}\\
Zero entries in the mass matrice lead to relations between mixing angles and mass 
ratios. In the framework of SUSY-GUTs, Roberts, Ramond  and Ross~\cite{RRR}
found five possible sets of symmetric mass matrices with more than three
texture zeros which are
consistent with the charge fermions phenomenology. 
Those can be extended into the leptonic sector using the good SO(10) ($E_6$) GUT
relations:
 
a) $Y_\tau (M_{GUT}) = Y_d (M_{GUT})$  \quad (induced via one $H_{\bf 10}$) \\
using the renormalization group equations to run  the relation down, 
one obtains the ``observed'' 
$$
3Y_\tau (1 GeV)  \simeq Y_b (1 GeV) \quad .
$$ 
b)  Georgi-Jarlskog~\cite{gj} relations (induced via $H_{\bf 126}$):
$$
Y_s (M_{GUT})  =  \frac{1}{3} Y_\mu (M_{GUT})\quad ; \quad Y_d (M_{GUT})
=3Y_e (M_{GUT})
$$
i.e. \qquad  $det M_\ell  (M_{GUT})  = det M_d (M_{GUT})$\quad .
 
c) similar relations between $M_u$ and $M_\nu^D$ are possible but GUT relations
do not say what $M_\nu^R$  is.
 
{\it How can we predict  $M_\nu^R$ ?} 
 
{\bf Several examples and ideas:}\\
(i) {\it The simplest possibility}:\\
To use a symmetry which forces {\em all} Yukawa 
matrices (including $M_\nu^R$) to have the same texture~\cite{ag1}.\\
This horizontal symmetry, on top of the SUSY-GUT gives additional relations between the
entries. In particular $M_\nu^D$ and $M_\nu^R$  are fixed in terms of the matrices of
the charged fermions. E.g. The symmetric Fritzsch texture~\cite{f} known  to give a 
large lepton-mixing~\cite{smir}.
 
One can use therefore
$$
|M_\nu^R| \simeq M_{GUT}
$$
to solve the SNP via vacuum oscillation of $\nu_1 - \nu_2$.\\ 
Problem: the model requires $m_t < 150 {\rm GeV}$  (can be avoided with small changes?)
 
(ii) {\it Minimality + Predictability\/}~\cite{dhr} \cite{ag2} \\
All matrix elements of the Yukawa matrices are due to one VEV only. The horizontal
 symmetry and the minimal Higgs structure fix $M_\nu^R$ in
terms of the parameters of the quark and charged leptons. 
Results: ``The standard scenario'': I.e. small letonic mixing angles which can solve the 
solar neutrino puzzle  via MSW. Also $\nu_3$ can be the hot dark matter.
\vfill\eject
(iii) {\it Asymmetric textures\/}\\
Most recent models for the quark mass (and mixing) use asymmetric mass matrices but
these were not yet applied for the neutrino-sector. An interesting exception is, however, 
models  of Babu and Barr~\cite{bar} with large leptonic mixing. In these models new 
vector like fields (which get
explicit large masses) are added. The latter have mixed mass terms with the light fermions in
such a way that the LH mixing angles of the quarks are small while the RH ones are large,
but the opposite is true for the lepton.
 
(iv) {\it Broken gauged Abelian symmetry } $U(1)_X$\\
Based on the idea of Froggat and Nielsen~\cite{fn} as applied to SUSY-GUTs~\cite{leurer}.
$U(1)_X$ dictates zero entries in the mass matrices by forbiding certain Yukawa
couplings. The general idea is that the heavy family acquires masses by direct coupling
to the light Higges. The light families obtain their masses when $U(1)_X$ is
spontaneously broken. This is done by giving a VEV to a chiral SM singlet field
$\theta$ with the charge $X_\theta = -1$. The GUT obtains then non-renormalizable
contribution like
$$
n_{ij} \overline{\Psi_i^c} \Psi_j H(\frac{\theta}{M})^{(Q_i + Q_j + Q_H)} 
$$   
where $Q_i$ are the $U(1)_X$ charges of the field and $M$ the scale of the
$U(1)_X$ invariant underlying theory (e.g. $M_{Planck}$).
 
Noting $\epsilon = \frac{<\theta>}{M} << 1$  and $Q_{ij} = 
Q_i + Q_j + Q_H > 0$, the Yukawa matrices have then the general form
$$
Y = \left(\begin{array}{ccc}
n_{11}e^{Q_{11}} & n_{12}e^{Q_{12}} & n_{13}e^{Q_{13}} \\
n_{21}e^{Q_{21}} & n_{22}e^{Q_{22}} & n_{23}e^{Q_{23}} \\
n_{31}e^{Q_{31}} & n_{32}e^{Q_{32}} & n_{33}e^{Q_{33}} \\
\end{array}\right)
$$

Assuming $n_{ij} \sim O(1)$ one can choose the $Q_{ij}$  such that a fit to the
quark masses and mixing is obtained. At this point GUT relations to obtain
$M_\ell$ and $ M_\nu ^D$  can be used. The models~\cite{u1} differ in details
especially for  $ M_\nu ^R$.
The resulting neutrino properties give  in general the ``Standard Scenario''.
This is by far the most popular type of models for the leptonic sector. Its nice property
is that it ``explains'' the observed hierarchy in masses and mixing
angles of the charged fermions. It is however, NOT SO PREDICTIVE FOR
THE NEUTRINO-SECTOR.
This is because of the
$ n_{ij} \sim O(1)$ unknown factors. $M_\nu^{light}$ in the see-saw expression
 eq.(1) is a product of three matrices, hence THERE IS AN UNKNOWN
 CORRECTION OF 
$[O(1)]^3$!\ \footnote{Note, that the $n_{ij}$ can be quite large like the Georgi-
Jarlskog $n_{22} = -3 $ factor.} Also, the charges $Q_{ij}$ are fixed by hand.
 
(v) {\it Correlated zeros in $M_\nu^D$ and $M_\nu^R$}.\\
This is a way  suggested by Bin\' etruy {\it et al}~\cite{bin} to solve the last problem
in case (iv). The correlation induces a natural mass degeneracy (independent of $n_{ij}$)
 and large mixing angles. Its breaking  leads  to only small $O(1)$ deviation from the full 
degeneracy (and not  $[O(1)]^3$ as in the previous case). They give several explicit 
examples and compare them with the neutrino-anomalies. The idea is very nice but the 
models use  in the present form many ad hoc requirements.
\vfill\eject
(vi) {\it  Non-Abelian family symmetries}\\
are the natural way to avoid $O(1)$ factors. One can fix at least part of the coefficients of the
non-renormalizable contributions in terms of Clebsch-Gordans of the non-Abelian 
group~\cite{dhr}.
 
Another advantage here is that one can account for the asymmetry between the heavy
family and the light ones. This is done by putting the families in a
 {\bf 1}~+~{\bf 2}
representation under U(2)~\cite{u2} or $S(3)^3$~\cite{s3}.
 
An interesting new model based on the family group $\Delta(48) \times U(1)$ is due
to Chou and Wu~\cite{cw}. They added a sterile neutrino and get:
$$
m_2 \approx m_3 >> m_1  \quad .
$$  
(vii) {\it Models with sterile neutrinos}.\\
The only case which can account for all the neutrino-anomalies. Needs a SM singlet fermion
which is light $\sim M_\nu$ and mixed with the light neutrinos. Questions: Why is it so
light? How is it mixed? why only one sterile neutrino? A possible solution~\cite{sn}
is to add new
matter and use again a kind of see-saw. A U(1) symmetry is then needed to protect the 
sterile neutrino mass. It was pointed out by Ma~\cite{ma} that $E_6$ GUT (in contrast with
SO(10)) has all the right ingredients for models with a sterile neutrino: I.e. a SM singlet
fermions and heavy matter in the {\bf 27} representation, as well as
additional U(1) symmetries. 

(viii) {\it GUT models (without low energy SUSY)} \footnote{SUSY is probably
required to include Gravity but is broken at $M_{Planck}$ in this case.}.\\
One uses in those models an intermediate Pati-Salam like symmetry at
$M_I \approx~10^{12} GeV$ to get Gauge and Yukawa unification with $M_{GUT} > 
5 \times 10^{15} {\rm GeV}$ so that problems with too fast proton decay are avoided. $M_I$ is
then a natural scale for $M_{\nu_R}$.
In the corresponding models:

a) Achiman and  Lukas~\cite{al} generate the mixing by GUT radiative
corrections induced via $M_\nu^R$ . This gives large leptonic
mixing.  The consistency of the Renormalization
Group Equations for this model require $m_t = 175 {\rm GeV}$~\cite{wicke}.
 
b) Babu and Mohapatra\cite{bm} show that mixing can be generated
via non-renormalizable contributions. They get however small
neutrino-mixing.
 
c) Lee and Mohapatra~\cite{lm} - use see-saw $II^{ed}$  art i.e.
$(M_\nu^{light})_{11} \ne 0$\ . This gives approximately degenerate neutrinos.
 
d) Buccella {\it et al}~\cite{buccella} suggested recently a new version based on
SO(10).
 
{\bf Concluding remarks}
 
Recent models incline toward using  asymmetric mass matrices (more freedom)
as well as giving the neutrinos degenerate masses, which go usually with large leptonic
mixing angles.
 
GUTs are a natural framework for the study  of the fermionic masses and mixing. And this
is true in particular for the neutrinos as they have the large scale needed for the see-saw
mechanism. 
 
Low energy theories for the neutrino-masses~\cite{val} cannot use the advantages
of GUTs. They need a lot of ad-hoc assumptions and new ``light'' particles. These lead 
obviously to many new phenomena one can look experimentally for.
 
We have seen that different models can give similar results. To distinguish  between models 
one needs additional information from outside the neutrino sector. E.g.  effects due
to RH mixing can play this role.  Recall that mass matrices are diagonalized 
via a bi-unitary transformation 
$$
M_{diag.} = U_R MU_L^\dag \quad .
$$
If $M$ is not symmetric (or hermitian)  $U_R \ne U_L$\ .
$U_R$ is not relevant for the SM as the RH components of the fermions are singlets. The RH 
rotation play however an important role in GUTs. E.g. in the proton decay\cite{ab}  or in 
$\nu_R$ decays which lead to leptogenesis.

\end{document}